\documentclass[12pt]{book}

\usepackage[dvips]{graphicx,color}
\usepackage{makeidx,observe,cosmology}

\makeauthorindex
\makeindex

\BookTitle{New Trends in Theoretical and Observational Cosmology}
\CopyRight{\copyright 2001 by Universal Academy Press, Inc.}

\def\kms{\ifmmode {\,\rm km \, s^{-1}}
\else {$\rm km \, s^{-1}$}\fi}
\def\Mpc{\ifmmode {\, h^{-1} \, {\rm Mpc}}
\else {$h^{-1}\,$ Mpc}\fi}

\def\s8{{\sigma_8}}

\def\ltsima{$\; \buildrel < \over \sim \;$}
\def\simlt{\lower.5ex\hbox{\ltsima}} 
\def\gtsima{$\; \buildrel > \over \sim \;$} 
\def\simgt{\lower.5ex\hbox{\gtsima}}

\def\omegam{{\Omega_{\rm m}}}

\def\omegab{{\Omega_{\rm b}}}

\def\omegal{{\Omega_\Lambda}}
 
\def\omegabh2{{\omegab h^2}}

\def\s8m{{\sigma_{8{\rm m}}}}
\def\s8g{{\sigma_{8{\rm g}}}}

\def\hompc{\ifmmode {\,h\,\rm Mpc^{-1}}
\else {$h^{-1}$~Mpc}\fi}
\def\m@th{\mathsurround=0pt }
\def\eqalign#1{\null\,\vcenter{\openup1\jot \m@th
 \ialign{\strut\hfil$\displaystyle{##}$&$\displaystyle{{}##}$\hfil
 \crcr#1\crcr}}\,}

\begin{document}

\BookTitle{\itshape New Trends in Theoretical and Observational Cosmology}
\CopyRight{\copyright 2001 by Universal Academy Press, Inc.}
\pagenumbering{arabic}

\chapter{
The 2dF Galaxy Redshift Survey:
Cosmological Parameters and Galaxy Biasing}

\author{%
Ofer LAHAV \& the 2dFGRS team\\
{\it Institute of Astronomy, University of Cambridge, Madingley Road, Cambridge
CB3 0HA, UK}}
%
%
\AuthorContents{O. Lahav} 

\AuthorIndex{Lahav}{O.}

\section*{Abstract}

The 2dF Galaxy Redshift Survey (2dFGRS) has already measured over
220,000 redshifts of nearby ($z\sim0.1$) galaxies.  
It allows us to
estimate fundamental cosmological parameters and to subdivide the
survey into subsets according to the galaxy intrinsic properties. The
large-scale structure analysis of the survey has already yielded
several significant results: (i) the shape of the power spectrum of
fluctuations is consistent with the $\Lambda$-Cold Dark Matter
concordance model; (ii) from joint analysis of the 2dFGRS data with
the Cosmic Microwave Background
anisotropies the linear-theory rms mass fluctuations is
$\sigma_{8{\rm m}} \approx 0.73$, lower than  the
COBE-alone normalization and previous estimates from 
cluster abundance and cosmic shear;
(iii) The biasing parameter of
bright galaxies on scales  $ \simgt  10 \Mpc$ is nearly unity; and
(iv) on scales smaller than $ \simlt 10 \Mpc$ red galaxies are more
strongly clustered than blue galaxies.


%

\section{Introduction}

Multifibre technology now allows us to measure  redshifts
of millions of galaxies. 
The Anglo-Australian 2
degree Field Galaxy Redshift Survey\footnote{The 2dFGRS Team comprises:
      I.J. Baldry, C.M. Baugh, J. Bland-Hawthorn, T.J. Bridges, R.D. Cannon, 
      S. Cole,
      C.A. Collins,  
      M. Colless,
      W.J. Couch, N.G.J. Cross, G.B. Dalton, R. DePropris, S.P. Driver,
      G. Efstathiou, R.S. Ellis, C.S. Frenk, K. Glazebrook, E. Hawkins, 
      C.A. Jackson,
      O. Lahav, I.J. Lewis, S.L. Lumsden, S. Maddox, 
      D.S. Madgwick, S. Moody, P. Norberg, J.A. Peacock, B.A. Peterson,
      W. Sutherland, K. Taylor. 
For more details on the survey see http://www.mso.anu.edu.au/2dFGRS/}
(2dFGRS)  
has already 
measured redshifts for 220,000 galaxies
selected from the APM catalogue  
(as of March 2002).  The median redshift of the
2dFGRS is ${\bar z} \sim 0.1$.  
It aims to acquire a complete sample of $\sim$250,000 galaxy spectra, 
down to an
extinction corrected magnitude limit of $b_J<19.45$ (Colless et al. 2001). 
A sample of this size allows large-scale structure statistics
to be measured with very small random errors. 
In this review we summarize some recent results 
from the 2dFGRS on clustering and galaxy biasing.


\section{The Power spectrum of 2dF Galaxies}

  An initial estimate of the convolved, redshift-space power spectrum of the
 2dFGRS has already been determined (Percival et al. 2001; hereafter P01)
 for a sample of 160,000 redshifts. 
 On scales $0.02<k<0.15 \hompc$, the data are
 robust and the shape of the power spectrum is not affected by
 redshift-space or non-linear effects, though the amplitude
 is increased by redshift-space distortions.
 P01 and Efstathiou
 et al. (2002; hereafter E02) have mainly compared the {\it shape} 
 of the 2dFGRS and CMB
 power spectra, and concluded that they are consistent with each other
(see also Tegmark, Hamilton \& Xu 2001), within the $\Lambda$-CDM
 framework.

Lahav et al. (2001) have estimated the {\it amplitudes} 
of the linear-theory rms fluctuations in mass $\sigma_{8{\rm m}}$ 
and in galaxies $\sigma_{8{\rm g}}$.
More precisely,  consider the ratio
of galaxy to matter power spectra, and use the ratio of these to
define the bias parameter:
\begin{equation}
b^2 \equiv {\sigma^2_{8{\rm g}} \over \sigma^2_{8{\rm m}} }
\equiv  {P_{\rm {gg}}(k)\over P_{\rm {mm}}(k)}
\end{equation}
On  scales of
$0.02 < k < 0.15 \hompc$ 
the fluctuations are close
to the linear regime, and there are good reasons (e.g. Benson et al. 2000)
to expect that $b$ should tend to a constant.
Here we do not test the assumption
that the biasing is scale-independent,
but we do allow it to be  function of luminosity
and redshift. 
Another necessary complication is that we need to distinguish
between the apparent values 
of $\sigma_{8{\rm g}}$ as measured in redshift space
($\sigma_{8{\rm g}}^S$) and the real-space value that would be measured in the
absence of redshift-space distortions ($\sigma_{8{\rm g}}^R$). It is the latter
value that is required in order to estimate the bias.
We emphasize  that here  $\sigma_{8{\rm g}}$ is
the linear-theory normalization, not the observed non-linear 
$\sigma_{8{\rm g}\rm NL}$. 
For example, the 2dFGRS correlation function of Norberg et al.
(2001a)
can be translated to a non-linear
$\sigma_{8{\rm g}\rm NL}^R(L_*) = 0.87 \pm 0.07$,
at an effective redshift of approximately 0.07.
In practice, nonlinear corrections to $\sigma_8$ are expected to
be relatively small for CDM-like spectra.

\begin{figure}[t]
  \begin{center}
    \includegraphics[height=15pc]{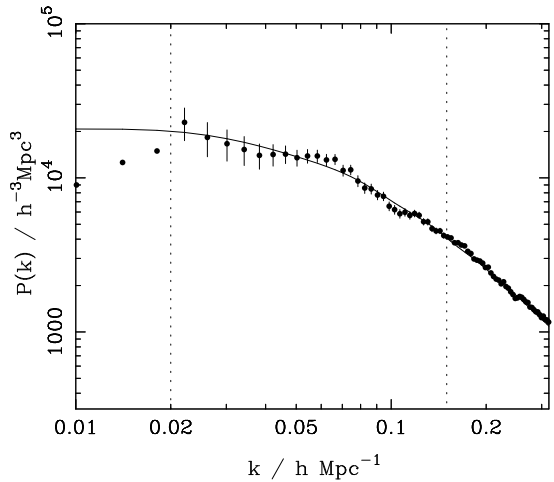}
  \end{center}
  \caption{The observed (i.e. convolved with the window function) 
2dFGRS power spectrum (as derived in Percival et al. 2001). 
The  solid line shows  a linear theory $\Lambda$-CDM fit 
(convolved with the window function)
with $\omegam h = 0.2, \omegab/\omegam = 0.15, h=0.7, n=1$ 
and best-fitting $\sigma^S_{8{\rm g}}(z_s,L_s) = 0.94$.
Only the range $ 0.02 < k < 0.15 \hompc$ is used 
at the present analysis
(roughly corresponding to CMB harmonics $200 <  \ell < 1500$
in a flat $\omegam = 0.3$ Universe).
The good fit of the linear theory power spectrum 
at $k > 0.15 \hompc$ is due to a conspiracy between the 
non-linear growth and finger-of-god smearing.}
\end{figure}

The 2dFGRS power spectrum (Fig. 1) is fitted in P01 
over the above range in $k$, 
assuming scale-invariant primordial 
fluctuations and a $\Lambda$-CDM cosmology, for 
four free parameters: $\omegam h$, $\omegab/\omegam$, $h$  
and the redshift space $\sigma^S_{8{\rm g}}$.
Assuming a 
Gaussian prior
on the Hubble constant $h=0.7\pm0.07$ (based on Freedman et al. 2001) 
 the shape of the recovered spectrum
within the above $k$-range
was used to yield 68 per cent confidence limits on the 
shape parameter 
$\omegam h=0.20 \pm 0.03$, and
the baryon fraction $\omegab/\omegam=0.15 \pm 0.07$,
in accordance with the popular `concordance' model.
For fixed 
`concordance model' parameters
$n=1, \omegam = 1 - \omegal = 0.3$, $\omega_{\rm b} = 0.02$
and a Hubble constant $h=0.70$, 
we find that the amplitude of 2dFGRS galaxies
in redshift space is $\sigma_{8{\rm g}}^S (L_s,z_s) \approx 0.94$.
As shown in P01, the likelihood analysis
gives a second (non-standard) solution, 
with   $\omegam h \sim 0.6$, and
the baryon fraction $\omegab/\omegam=0.4$,
which generates baryonic `wiggles'. We ignore this case at the 
present analysis.
We also note that even if there are features in the primordial 
power spectrum, they would get washed out by the 2dFGRS 
window function (Elgaroy, Gramann \& Lahav 2002).

In reality, the effective redshift for the P01 analysis is not zero,
but $z_s \sim 0.17$.  This is higher than the median redshift of
2dFGRS ($z_m \sim 0.11$) due to the weighting scheme used in
estimating the power spectrum.  Similarly, $L_s\simeq 1.9L_*$, rather
than the $L_s\simeq L_*$ that would apply for a flux-limited sample.
The effects of redshift-space distortion, evolution of biasing
and luminosity bias 
on the determination of $b$
are quite
significant, at the level of $ \sim 15 $ per cent each.

\section {Combining 2\lowercase{d}FGRS \& CMB} 

A simultaneous analysis of the constraints placed on cosmological
parameters by different kinds of data is essential because each probe
(e.g. CMB, SNe Ia, redshift surveys, cluster abundance, and peculiar
velocities) typically constrains a different combination of
parameters (e.g. Bahcall et al. 1999; Bridle et al. 2001a; E02).
A particular case of joint analysis 
is that of galaxy redshift surveys and the 
CMB.
While the CMB probes the fluctuations in matter, 
the  galaxy redshift surveys measure  the perturbations 
in the light distribution  of  particular tracer
(e.g. galaxies of  certain type). 
Therefore, for a fixed set of cosmological  parameters, 
a combination of the two can tell us about the way galaxies
are `biased' relative to the mass fluctuations (e.g. Webster et al. 1998).

The CMB fluctuations are commonly represented by the 
spherical harmonics $C_{\ell}$.
The connection between the harmonic $\ell$ and $k$ is roughly
\begin{equation}
\ell \simeq k\, {\frac{2c}{H_0 \,\Omega_{\rm m}^{0.4}}}\;  
\end{equation} 
for a flat Universe.
For $\omegam =0.3$ 
the 2dFGRS  range $0.02 < k < 0.15 \hompc$
corresponds 
approximately to $ 200 < \ell < 1500$, which is 
well covered by the recent CMB experiments.

A well-known problem in estimating cosmological parameters is the 
degeneracy of parameters,
and the choice of free parameters.
Here we assume 
a flat Universe (i.e. zero curvature), and no tensor component
in the CMB
(for discussion of the degeneracy with respect to these
parameters see E02).
We consider five free parameters: the matter 
density parameter $\omegam$,
the linear-theory amplitude of the mass fluctuations $\sigma_{8{\rm m}}$, 
the present-epoch linear biasing parameter 
$b(L_s,z=0)$ (for the survey effective luminosity $L_s \simeq 1.9 L_*$), the 
Hubble constant  $ h \equiv H_0/(100 \kms)$, and
the baryon density parameter $\omega_{\rm b} \equiv \omegab h^2$.
As we are mainly interested in 
combinations of $\sigma_{8{\rm m}}$, $b$ and $\omegam$, 
we shall marginalize
over the remaining  parameters.
We also check the robustness of the 
results to some `extra parameters',  
 the optical depth $\tau$ due to reionization 
and the primordial spectral index $n$.


The latest CMB measurements 
from Boomerang (Netterfield et al.\ 2001, de Bernardis et al.\
2002), Maxima (Lee et al. 2001; Stomper et al.\ 2001) 
and DASI (Halverson et al. 2002; Pryke et al.\ 2002) 
suggest three acoustic peaks.  Parameter fitting
to a $\Lambda$-CDM model indicates consistency between the different
experiments, and a best-fit Universe with zero curvature, and an
initial spectrum with spectral index $n  \simeq 1$
(e.g. Wang et al.\ 2001, E02
and references therein).  Unlike the earlier Boomerang and Maxima
results, the new data also show that the baryon contribution is
consistent with the Big Bang Nucleosynthesis value $\omega_{\rm b} 
\simeq 0.02$ (O'Meara et al.\ 2001).
We have used a  compilation of  COBE, Boomerang, Maxima and DASI
data, after marginalization over calibration errors
(Bridle et al. 2001b).

\begin{figure}[t]
  \begin{center}
    \includegraphics[height=13pc]{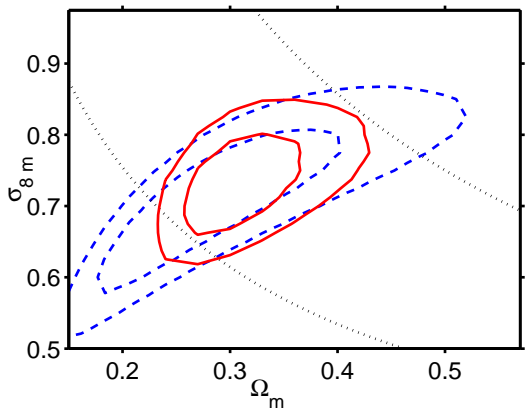}
  \end{center}
  \caption{The likelihood function of CMB alone  (dashed lines) in terms
of the mass fluctuation amplitude $\sigma_{8{\rm m}}$ and the present 
epoch $\omegam$.
The marginalization over the Hubble constant is done 
with a Gaussian centred at $h=0.7$ and standard deviation of 0.07.
Other parameters are held fixed ($n=1, \omega_{\rm b} = 0.02, \tau=0.0$).
The contours are for (two-parameter) 68 per cent and 95 per cent confidence intervals.
The solid lines show the contours (68 per cent and  95 per cent) 
for the joint 2dFGRS+CMB analysis,
after marginalization over $h$, $b(L_s,0)$ and $\omega_{\rm b}$.
Other parameters are held fixed ($n=1, \tau=0.0$).
Note that the contours of 2dFGRS+CMB are much tighter than 
when using CMB alone. 
Two recent extreme cluster abundance determinations 
are overlayed as  the upper
dotted line (Pierpaoli et al. 2001)
and the lower dotted line (Viana et al. 2002).
From Lahav et al. (2001).
}
\label{cmb_s8_om}
\end{figure}

When combining 2dFGRS and CMB data
the parameterization for the log-likelihoods is
then:
\begin{equation}
\eqalign{
{\ln {\mathcal L}}_{\rm tot} &= \ln {\mathcal L}_{\rm 2dFGRS} [\omegam, h, 
\omega_{\rm b}, \sigma_{8{\rm m}}, b(L_s,0)] \cr
& + \ln  {\mathcal L}_{\rm CMB} [\omegam, h, \omega_{\rm b}, \sigma_{8{\rm m}}],
}
\end{equation}
where
${\mathcal L}_{\rm 2dFGRS}$ and
${\mathcal L}_{\rm CMB}$ 
are the likelihood functions for 2dFGRS and the CMB.

The 2dFGRS likelihood function takes into account the redshift-space distortions, 
an epoch-independent galaxy clustering 
biasing scheme, and the redshift evolution of $\omegam(z)$.
Other parameters are held fixed 
($n=1, \tau=0$).

Fig. ~\ref{cmb_s8_om} (solid lines) shows 
the 2dFGRS+CMB likelihood as a function of
$(\omegam, \sigma_{8{\rm m}})$, after marginalization over
$h, b(L_s,0)$ and $\omega_{\rm b}$.
The peak of the distribution is consistent with the result 
for the CMB alone (shown 
by the dashed lines in Fig. ~\ref{cmb_s8_om}),
but we see that the contours are tighter due
to the addition of the 2dFGRS data. 
Further marginalization over $\omegam$  
gives  $\sigma_{8{\rm m}} = 0.73 \pm 0.05$.

To study the biasing parameter
we marginalize the 2dFGRS likelihood 
over $h, \omega_{\rm b}$, 
$\sigma_{8{\rm m}}$ and $\omegam$
(with fixed  parameters are held fixed $n=1, \tau=0$)
and we get 
$b(L_s,0)=1.10 \pm 0.08$ (1-sigma).

To translate the biasing parameter from $L_s$ to
e.g. $L_*$ galaxies one can either assume
(somewhat ad-hoc) no luminosity segregation on large 
scales, or divide by the factor 1.14 (Norberg et al. 2001a) that applies
on small scales.
We also tested sensitivity to the spectral index $n$ 
and the optical depth $\tau$.
Overall, our results can be described by the following formula:
\begin{equation}
b(L_*,z=0) = (0.96 \pm 0.08)\, \exp[-\tau + 0.5(n-1)].
\end{equation}


\section {Comparison with other measurements }

\subsection{Other estimates of 2dFGRS amplitude of fluctuations}

An independent  measurement from 2dFGRS comes from 
redshift-space distortions on scales $ \simlt 10 \Mpc$ (Peacock et al. 2001).
This gives 
$\beta(L_s,z_s) = 0.43 \pm 0.07$.
Using the full likelihood function 
in the $(b, \omegam$) plane we derive a slightly 
larger (but consistent) value,
$\beta(L_s,z_s) \simeq  0.48 \pm 0.06$. 

A study of the bi-spectrum of the 2dFGRS (Verde et al. 2001) 
on smaller scales ($0.1 < k < 0.5 \hompc$) sets  constraints 
on deviations from linear biasing, and it gives a best-fit solution 
consistent with  linear biasing of unity.
The  agreement with the 2dF+CMB result  is impressive,
given that the methods used are entirely different.
In fact, by matching the two results one can get constraints 
on e.g.  the optical depth $\tau  \simlt  0.2 $.

\subsection {Comparison with other independent measurements}

Our derived values for
$\sigma^R_{8{\rm g}}(L_*, z=0)$ is 
in accord with 
with the values derived from 
the early Sloan Digital Sky Survey (SDSS) 
by Szalay et al. (2001).


Cluster abundance is a popular method for constraining $\sigma_{8{\rm
m}}$ and $\omegam$ on scales of $\sim 10 \Mpc$.  Four recent analyses
span a wide range of values, but interestingly they are all orthogonal
to our CMB and 2dF constraints (Fig. 3).  For example, for
$\omegam=0.3$. Pierpaoli, Scott \& White (2001), Seljak (2001),
Reiprich \& Boehringer (2002), and Viana, Nichol \& Liddle (2002)
found: $\sigma_{8{\rm m}} \simeq 1.02; 0.75; 0.68; 0.61$ respectively
(with typical errors of 10 per cent).  The discrepancy between the
different estimates is in part due to differences in the assumed
mass-temperature relation.  The cluster physics still needs to be
better understood before we can conclude which of the above results is
more plausible.  We see in Fig. 2 that the lower cluster abundance
results are actually in good agreement with our value from the
2dFGRS+CMB, $\sigma_{8{\rm m}} \simeq 0.73 \pm 0.05$.


The measurements of weak gravitational lensing (cosmic shear) 
are sensitive to the amplitude of the matter
power spectrum on mildly non-linear scales. 
For example, for $\omegam = 0.3$  
Bacon et al. (2002)  and Van Waerbeke et al. (2002) find
$\sigma_{8{\rm m}} \simeq  0.97$ 
(with errors of about 20 per cent).
These estimates  are higher than the $\sigma_{8{\rm m}}$  value
that we obtain from 2dFGRS+CMB, but note the large error bars
in this recently developed  method.    

\section {Clustering per spectral type}

Although biasing was commonly neglected until the early 1980s,
it has become evident that on scales $ \simlt 10 \Mpc$  
different galaxy populations exhibit 
different clustering amplitudes, the so-called
morphology-density relation (e.g. Dressler 1980; Hermit et al. 1996). 
Biasing on small scales is also predicted in the simulations
of hierarchical clustering from CDM initial conditions 
(e.g. Benson et al. 2000).
It is important therefore to pay attention to the scale 
on which biasing operates.

\begin{figure}[t]
  \begin{center}
    \includegraphics[height=13pc]{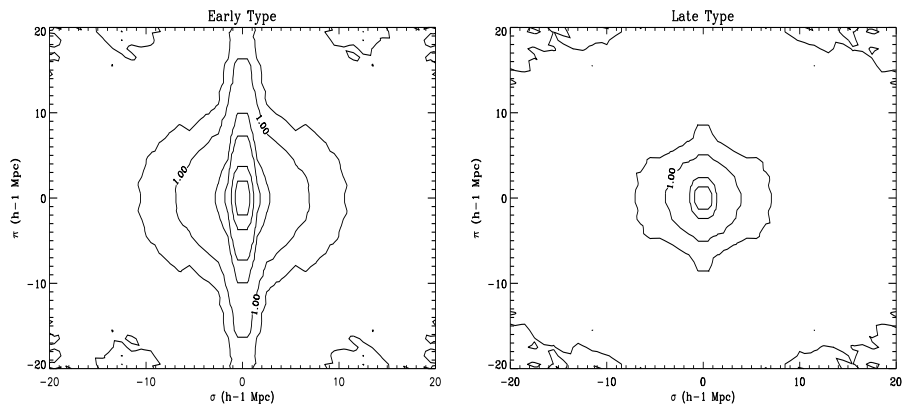}
  \end{center}
  \caption{The two point correlation function $\xi(\sigma,\pi)$ plotted for
   passively (left) and actively (right) star-forming galaxies.  
The contour levels are 15,10,5,3,1 (labelled),0.5,0.
From Madgwick \& Lahav (2001) and Madgwick, Hawkins et al. (in preparation).}
\label{xiplot}
\end{figure}

Madgwick et al. (2002) have utilized the method of Principal 
Component Analysis (PCA) to compress each galaxy spectrum
into one quantity, $\eta \approx 0.5\;pc_1 + pc_2$.
Qualitatively, $\eta$ is an indicator of the current star formation
in a galaxy.
This allows us to divide the 2dFGRS into $\eta$-types, 
and to study e.g. luminosity functions  
and clustering per type.

Norberg et al. (2002) found that 
for $L_*$ galaxies, the real space
correlation function amplitude 
of $\eta$ early-type galaxies is $\sim 50 \% $ 
higher than that of late-type galaxies. 
Fig 3 (from Madgwick \& Lahav 2001) shows the 
redshift space correlation 
function in terms of the line-of-sight 
and perpendicular to the line-of-sight separation $\xi(\sigma,\pi)$. 
In this plot we show the correlation function calculated
from the most passively (`red') and actively (`blue') star-forming 
galaxies.
It can be seen that the clustering properties of the
two samples are quite distinct on scales $\simlt 10 \Mpc$.  
The `red' galaxies 
display a prominent `finger-of-god' effect and also have a higher overall
normalization than the `blue' galaxies.
This is a manifestation of the well-known morphology-density relation.
This diagram allows us to determine the combination of the mass density 
and biasing parameter $\beta = \Omega_m^{0.6}/b$ 
(e.g. Peacock et al. 2001; Hawkins et al., in preparation).

\section {Discussion} 

We have combined  the latest 2dFGRS and CMB data.
The first main  result of this joint analysis is 
the normalization of the mass fluctuations, 
$\sigma_{8{\rm m}} = 0.73 \pm 0.05$.
This normalization is lower than the COBE normalization and 
previous estimates from cluster abundance, but it is
actually in agreement with recently revised cluster abundance normalization.
The results from cosmic shear are still somewhat higher,
but with larger error bars. 
 
The second result is for the biasing parameter for
optically-selected $L_s$ galaxies,
$b(L_s,0) = 1.10 \pm 0.08$, which
is consistent with  no biasing (`light traces mass') on scales
of tens of Mpc.
When translated to $L_*$ via a correction valid for  small 
scales we get a slight anti-bias,
$b(L_*,0) \simeq 0.96$.
Our result of linear biasing of unity on scales ($ \simgt  10 \Mpc$)
is actually in agreement with predictions of simulations
(e.g. Blanton et al. 2000 Benson et al. 2000; Somerville et al. 2001).
It was also demonstrated  by Fry (1996)
that even if biasing was larger than unity at high redshift, 
it would converge towards unity at late epochs.
Yet on scales $\simlt 10 \Mpc$ 
different galaxy populations have different clustering amplitudes. 

It may well be that in the future the
cosmological parameters
will be fixed by CMB, SNe etc.
Then, for fixed reasonable cosmological parameters,
one can use redshift surveys to study biasing, evolution, etc.
The above analysis is a modest illustration of this approach.
Future work along these lines will include
exploring non-linear biasing models 
(e.g. Dekel \& Lahav 1999; Sigad, Branchini \& Dekel  2001; Verde et al. 2001)
per spectral type or colour
(Madgwick et al 2001; Norberg et al. 2002; 
Madgwick, Hawkins et al., in preparation; Zehavi et al. 2001)
and the detailed variation of other galaxy properties with 
local mass density.

Overall, the results from 2dFGRS fit well
into the `concordance' model which has emerged 
from various cosmological data sets.
The $\Lambda$-CDM model
with comparable amounts of dark matter and dark energy is rather esoteric,
but it is remarkable that different measurements
converge to the `concordance model' with parameters: 

\begin{itemize}
\item $\Omega_k  \approx 0$, 
\item $\Omega_m =1 - \Omega_\Lambda \approx 0.3$, 
\item $\Omega_b h^2 \approx 0.02$,
\item $h \approx 0.7$
\item the age of the Universe $t_0 \approx 14 $ Gyr, 
\item  the spectral index $n \approx 1$,
\item $\sigma_{8{\rm m}} \approx 0.75$.

Perhaps the least accurate estimates on that list are for $\omegam$ 
and $\sigma_{8{\rm m}}$.

\end{itemize}

\bigskip

While phenomenologically the $\Lambda$-CDM
model has been successful in fitting a wide range
of cosmological data, there are some open questions:

\begin{itemize}

\item
Both components of the model, $\Lambda$ and CDM, 
have not been directly measured.
Are they `real' entities or just `epicycles' ?
\item
Why is  $\Omega_m \sim \Omega_\Lambda$ at the present-epoch ?
Do we need to introduce  a new physics 
or invoke the Anthropic Principle to explain it ?
\item
There are still open problems in 
$\Lambda$-CDM on the small scales
e.g. galaxy profiles and satellites.
\item 
The age of the Universe is uncomfortably close to some estimates for the 
age of the Globular Clusters, when  their epoch of 
formation is also taken into account (Gnedin, Lahav \& Rees 2001). 
\item  
Could other (yet unknown) models fit the data equally well ?
\item
Where does the field go from here ?
Would the activity focus on refinement of 
the cosmological parameters within $\Lambda$-CDM, 
or on introducing entirely new paradigms   ?
\end{itemize} 

These issues will no doubt 
be revisited soon with larger and more accurate data sets.
We will soon be able to map
the fluctuations with scale and epoch, and to analyze jointly
redshift surveys 
(e.g. 2dF, SDSS) and
CMB (e.g. MAP, Planck) data. 
These high quality data sets 
will allow us to study a wider range of models and parameters.

\bigskip
\bigskip

\section*{ACKNOWLEDGMENTS} 

I thank Sarah Bridle, Darren Madgwick and
the 2dFGRS team members for their contribution 
to the work described here.
The 2dF Galaxy Redshift Survey was made possible through
the dedicated efforts of the staff of the Anglo-Australian Observatory,
both in creating the 2dF instrument and in supporting it on the telescope.
I also thank the conference organizers for the hospitality in Tokyo.



\end{document}